\documentclass[reprint,superscriptaddress,nofootinbib,aps,prc]{revtex4-2}

\usepackage{graphicx}
\usepackage{dcolumn}
\usepackage{bm}
\usepackage{placeins}
\usepackage{lipsum}
\usepackage{feynmf}

\usepackage{amsmath}
\usepackage{amsfonts}
\usepackage{amssymb}
\usepackage{physics}

\usepackage{float}
\usepackage{hyperref}
\usepackage{color}

\newcommand{\Gth}{\Gamma_\mathrm{th}}
\newcommand{\Pth}{P_\mathrm{th}}
\newcommand{\Eth}{E_\mathrm{th}}

\newcommand{\fmmo}{\,\mathrm{fm}^{-1}}
\newcommand{\fmmt}{\,\mathrm{fm}^{-3}}
\newcommand{\MeV}{\,\mathrm{MeV}}

\begin{document}

\title{Neutron matter at finite temperature based on chiral effective field theory interactions}

\newcommand{\tud}{\affiliation{Technische Universit\"at Darmstadt, Department of Physics, 64289 Darmstadt, Germany}}
\newcommand{\emmi}{\affiliation{ExtreMe Matter Institute EMMI, GSI Helmholtzzentrum f\"ur Schwerionenforschung GmbH, 64291 Darmstadt, Germany}}
\newcommand{\mpi}{\affiliation{Max-Planck-Institut f\"ur Kernphysik, Saupfercheckweg 1, 69117 Heidelberg, Germany}}

\author{J.~Keller}
\tud
\emmi
\author{C.~Wellenhofer}
\tud
\emmi
\author{K.~Hebeler}
\tud
\emmi
\author{A.~Schwenk}
\tud
\emmi
\mpi

\begin{abstract} 

We study the equation of state of neutron matter at finite temperature based on two- and three-nucleon interactions derived within chiral effective field theory to next-to-next-to-next-to-leading order. The free energy, pressure, entropy, and internal energy are calculated using many-body perturbation theory including terms up to third order around the self-consistent Hartree-Fock solution. We include contributions from three-nucleon interactions without employing the normal-ordering approximation and provide theoretical uncertainty estimates based on an order-by-order analysis in the chiral expansion. Our results demonstrate that thermal effects can be captured remarkably well via a thermal index and a density-dependent effective mass. The presented framework provides the basis for studying the dense matter equation of state at general temperatures and proton fractions relevant for core-collapse supernovae and neutron star mergers.

\end{abstract}

\maketitle

\section{Introduction} 

Core-collapse supernovae and neutron star mergers are energetic events at the extremes. These fascinating and spectacular astrophysical phenomena probe strong interactions over a wide range of densities and temperatures.
Their multimessenger observations can be confronted with numerical simulations, which require information on the equation of state (EOS) as a key input.  
To date, all numerical EOS tables are based on phenomenological models, which make it difficult to assess strong interaction uncertainties.

Microscopic calculations based on modern nuclear interactions on the other hand make it possible to quantify theoretical uncertainties, at least up to nuclear densities. In this density regime the relevant baryonic degrees of freedom are neutrons and protons. Chiral effective field theory (EFT) provides a systematic low-energy expansion of the interactions between nucleons with a direct connection to the symmetries of quantum chromodynamics (QCD)~\cite{Epelbaum_et_al_2009, Machleidt_Entem_2011, Hammer2020}. Calculations based on such interactions at different orders in the expansion allow to estimate uncertainties due to omitted higher-order terms. Furthermore, three-nucleon (3N) interactions, known to be important for observables of atomic nuclei and matter~\cite{Hammer_et_al_2013,Hebeler_Holt_et_al_2015, Hebeler2020}, are determined consistently with two-nucleon (NN) interactions, and for neutrons they are predicted parameter-free to next-to-next-to-next-to-leading order (N$^3$LO)~\cite{Tews13N3LO,Krue13N3LOlong}.
In this work we employ a set of modern interactions that have been shown to predict masses of light and medium-mass nuclei as well as empirical saturation properties of symmetric nuclear matter at zero temperature in good agreement with empirical constraints~\cite{Hebeler_et_al2011,Drischler2019,Simo17SatFinNuc,Stro20limits}.

While nuclear matter at zero temperature has been investigated quite extensively based on chiral EFT interactions~\cite{HebelerSchwenk2010,Hebeler_et_al2011,Tews13N3LO,Holt13PPNP,Carb13nm,Hage14ccnm,Coraggio_et_al_2014,Lynn16QMC3N,Dris16asym,Dris16nmatt,Ekst17deltasat,Logo2019consistent, Drischler2019}, studies at finite temperature are less advanced. This is an unsatisfying situation as, e.g., recent core-collapse supernova simulations have demonstrated the importance of a proper treatment of finite-temperature effects in the EOS~\cite{Yasin_et_al_2020,Schneider:2019shi}. In neutron star merger simulations, thermal effects are sometimes approximated via a constant thermal index (see, e.g., Ref.~\cite{Bauswein:2010dn}). The availability of microscopic calculations over the full range of relevant temperatures would make such approximations obsolete, at least up to densities where nuclear interactions are applicable and reliable. For recent work that implements chiral EFT constraints into EOS functionals, with a focus on thermal effects, see Ref.~\cite{Huth:2020ozf}. Because these dense astrophysical environments tend to be neutron-rich, this paper focuses on finite-temperature calculations of neutron matter, but the calculational framework can be extended in a straightforward way to general proton fractions.

Nuclear matter at finite temperature has been studied with a range of many-body methods. In addition to calculations using the Brueckner-Hartree-Fock approach~\cite{Zuo_et_al_2004} and nominally variational calculations~\cite{FriedmanPandharipande1981}, the finite-temperature EOS has been calculated using many-body perturbation theory (MBPT)~\cite{Tolos:2007bh,Fiorilla_et_al_2012,PhysRevC.89.064009,PhysRevC.92.015801,PhysRevC.93.055802}, and nonperturbatively using the self-consistent Green’s function (SCGF) approach~\cite{Carb13nm, Carbone_et_al_2018, Carbone2020, Rios2020} and lattice EFT~\cite{Lu19matter}.

In this work, we take several steps towards improved finite-temperature MBPT calculations. For an efficient evaluation of individual diagrams, we represent NN and 3N interactions in a single-particle representation following the framework of Ref.~\cite{Drischler2019} and employ Monte Carlo sampling techniques to reliably compute the resulting high-dimensional MBPT phase space integrals in an efficient way. We treat 3N interactions explicitly, without employing  density-dependent two-body approximations (see, e.g., Refs.~\cite{HebelerSchwenk2010, Holt10ddnn}). Moreover, we include NN and 3N interactions through partial-wave decomposed matrix elements~\cite{Hebeler_et_al_2015,Hebeler2020}, which enables MBPT calculations for general nuclear forces.
To provide systematic uncertainty estimates, we employ a large set of chiral NN plus 3N interactions at different orders in the chiral expansion up to N$^3$LO. We take into account all contributions of NN interactions up to third-order in the MBPT expansion around the self-consistent Hartree-Fock (HF) solution, which implicitly includes contributions from anomalous diagrams at second and third order in MBPT. Finally, we provide a detailed analysis of thermal interaction effects and to which extent they can be approximated by a density-dependent effective mass and a thermal index, which is of interest for astrophysical applications.

This paper is organized as follows. In Sec.~\ref{sec2} we discuss the general MBPT framework at finite temperature, in particular the role of anomalous contributions and the simplifications when using a HF partitioning of the Hamiltonian. In Sec.~\ref{sec:results} we present results for various thermodynamic quantities and their uncertainties based on different nuclear interactions obtained from chiral EFT up to N$^3$LO. Moreover, we study the different contributions to thermal effects, and use the thermal index to extract the neutron effective mass. Finally, we summarize and conclude in Sec.~\ref{sec:conclusions}.

\section{Many-body framework}
\label{sec2}

With the development of chiral EFT interactions at low cutoff scales~\cite{Epelbaum_et_al_2009,Machleidt_Entem_2011} and renormalization group (RG) methods that allow to evolve interactions to lower resolution~\cite{Bogner:2009bt,Furnstahl:2013oba}, many-body perturbation theory (MBPT) becomes a viable and systematic approach to the nuclear many-body problem~\cite{BOGNER200559,Bogner:2009bt,DrischlerHoltWellenhofer2021,Tichai2020}. Here, in Sec.~\ref{sec21} we first provide a short review of the finite-temperature MBPT expansion around a general one-body Hamiltonian. In zero-temperature MBPT calculations it is common to use a HF reference state,  since this improves the many-body convergence compared to MBPT around the noninteracting Fermi gas. The generalization of HF-MBPT to finite temperatures involves some subtleties which we discuss in Sec.~\ref{sec22}.

\subsection{MBPT at finite temperature}
\label{sec21}

We determine the thermodynamic properties of neutron matter starting from the grand-canonical potential
\begin{align}
    \Omega(T,\mu) &= -\frac{1}{\beta} \ln Z\left(T, \mu\right)\,,
\end{align}
where ${Z(T, \mu) = \Tr\left(e^{-\beta \left(H - \mu N \right)}\right)}$ is the partition function of the system, with $T = 1/\beta$ the temperature, $N$ the particle number, and $\mu$ the chemical potential. The Hamiltonians $H$ considered in this work consist of the kinetic term ($H_0$) plus contributions from two- and three-nucleon interactions (see Sec.~\ref{sec:results}): 
\begin{align}
H = H_0 + V_{\rm NN} + V_{\rm 3N}\,.
\end{align}
Many-body perturbation theory offers the freedom to choose a specific partitioning of the Hamiltonian which defines the reference basis that is used for the perturbative expansion. The simplest choice consists in expanding $\Omega(T,\mu)$ about the noninteracting system with Hamiltonian $H_0$. However, usually the convergence of the expansion can be improved by choosing a more general partitioning of the form
\begin{align}\label{partitioning}
H =\left(H_0 + U \right) + \lambda \left(V_{NN} + V_{3N} - U \right)\,,
\end{align}
where the perturbation parameter $\lambda$ is eventually set to $\lambda=1$. Here, the operator $U$ corresponds to an effective single-particle potential, i.e.,\footnote{Here and in the following we use collective labels ${\alpha = (\bf{k}, \sigma)}$ for momentum $\bf{k}$ and spin projection $\sigma = \pm 1/2$, and the shorthand notation $\sum_\alpha f_\alpha = \sum_\sigma \int \frac{\mathrm{d}^3 {\bf k}}{(2\pi)^3} f(\bf{k}, \sigma)$.}
\begin{align}
    U = \sum_\alpha U_\alpha a^\dagger_\alpha a_\alpha\,,
\end{align}
with creation and annihilation operators $a^\dagger_\alpha$ and $a_\alpha$. The single-particle spectrum of the reference system is then given by
\begin{align}
    \varepsilon_\alpha = \frac{k^2}{2M} + U_\alpha\,.
\end{align}
The perturbation series of the grand-canonical potential is then obtained as (see, e.g. Refs.~\cite{FetterWalecka,NegeleOrland})
\begin{align}
    \Omega(T,\mu) &= \sum_{l=0}^{\infty} \lambda^l \Omega_l(T,\mu)\,,
\end{align}
where $\Omega_0(T,\mu) = -\frac{1}{\beta} \sum_\alpha \ln\left(1 + e^{-\beta \left(\varepsilon_\alpha - \mu \right)}\right)$ 
is the grand-canonical potential of the reference system.
The first-order contribution reads
\begin{align}
    \Omega_1(T,\mu) = &-\sum_\alpha n_\alpha U_\alpha\nonumber\\
    &+ \frac{1}{2}\sum_{\alpha\beta} n_\alpha n_\beta \bra{\alpha\beta}\mathcal{A}_{12} V_{\rm NN}\ket{\alpha\beta}\nonumber\\
    &+ \frac{1}{6}\sum_{\alpha\beta\gamma} n_\alpha n_\beta n_\gamma \bra{\alpha\beta\gamma}\mathcal{A}_{123}V_{\rm 3N}\ket{\alpha\beta\gamma}
    \,,\label{eq:hf-contribution}
\end{align}
where $\mathcal{A}_{12}$ and $\mathcal{A}_{123}$ are two- and three-particle antisymmetrizers, and the Fermi-Dirac distributions are given by
\begin{align}
    n_\alpha = \frac{1}{e^{\beta(\varepsilon_\alpha - \mu)} + 1}\,.
\end{align}
Equation~\eqref{eq:hf-contribution} matches the corresponding contribution 
to the ground-state energy in zero-temperature MBPT, 
with the Fermi-Dirac distributions replaced by $\theta(k_\text{F}-k)$.
This correspondence is lost at second order and beyond, where 
additional so-called anomalous contributions~\cite{KohnLuttinger1960,NegeleOrland,FetterWalecka} appear in finite-temperature MBPT.
The second-order contribution from two-body interactions is given by \begin{align}
  \Omega_{2}^\text{NN}(T,\mu) =   
  \Omega_{2,\text{normal}}^\text{NN}(T,\mu) + \Omega_{2,\text{anomalous}}^{\text{NN}}(T,\mu)\,,
\end{align}
where
\begin{align}
    \Omega_{2,\text{normal}}^\text{NN}(T,\mu) &= -\frac{1}{8}\sum_{\alpha\beta\gamma\delta} P_{\alpha\beta\gamma\delta}\, \abs{\bra{\alpha\beta}\mathcal{A}_{12}V_{\rm NN}\ket{\gamma\delta}}^2\label{eq:normal-diagram}\,,
\end{align}
with
\begin{align}
    P_{\alpha\beta\gamma\delta} &= \frac{n_\alpha n_\beta (1-n_\gamma) (1-n_\delta) - (1-n_\alpha)(1-n_\beta) n_\gamma n_\delta}
    {\varepsilon_\gamma + \varepsilon_\delta - \varepsilon_\alpha - \varepsilon_\beta}\,,
    \label{eq:phase2}
\end{align}
and    
\begin{align}
    \Omega_{2,\text{anomalous}}^{\text{NN}}(T,\mu) &= -\frac{1}{2T} \sum_{\beta} n_\beta \left(1 - n_\beta \right)
    \nonumber  \\ & \quad \times
    \left(\sum_\alpha n_\alpha \bra{\alpha\beta}\mathcal{A}_{12}V_{\rm NN}\ket{\alpha\beta} \right)^2\,.\label{eq:anomalous-diagram}
\end{align}
Furthermore, there are diagrams at second order that involve the effective one-body potential, which are shown in Fig. \ref{fig:anourmal_cancelation} below. Their analytical expressions are given by
\begin{align}
\Omega_{2}^{\text{NN-U}}(T,\mu) &= \frac{1}{2T} \sum_{\beta} n_\beta \left(1 - n_\beta \right)
\nonumber  \\ & \quad \times
\left(\sum_\alpha n_\alpha \bra{\alpha\beta}\mathcal{A}_{12}V_{\rm NN}\ket{\alpha\beta} \right)U_\beta\,,\label{eq:U1}\\
\Omega_{2}^{\text{U-NN}}(T,\mu) &= \frac{1}{2T} \sum_{\alpha} n_\alpha \left(1 - n_\alpha \right)
\nonumber  \\ & \quad \times
U_\alpha \left(\sum_\beta n_\beta \bra{\alpha\beta}\mathcal{A}_{12}V_{\rm NN}\ket{\alpha\beta} \right)\,,\label{eq:U2}\\
\Omega_{2}^{\text{U-U}}(T,\mu) &= -\frac{1}{2T} \sum_{\beta} n_\beta \left(1 - n_\beta \right) U_\beta^2\,.\label{eq:U3}
\end{align}
The anomalous contributions given by Eq.~\eqref{eq:anomalous-diagram}--\eqref{eq:U3} are absent in the zero-temperature formalism~\cite{KohnLuttinger1960,NegeleOrland,FetterWalecka}.
Note that the expression given by Eq.~\eqref{eq:phase2} has no poles at finite $T$. In the $T\rightarrow 0$ limit the two parts of the numerator in Eq.~\eqref{eq:phase2} separate into two identical contributions (with integrable poles at the integration boundary) whose sum matches the corresponding zero-temperature expression. The expressions for the second-order contributions involving three-nucleon interactions have similar features, and similar for contributions beyond second order. 

While we focus the discussion here mostly on the contributions from two-body interactions, in our calculations we include the complete set of second-order contributions. In particular, we include also the residual 3N contribution at second order~\cite{PhysRevC.94.034001,Drischler2019}. At third order we include all contributions that involve only NN interactions. Regarding the nonresidual third-order terms with 3N interactions, we have checked that their contribution in neutron matter is small compared to the corresponding diagrams containing only NN interactions. This is consistent with the findings of Ref.~\cite{Drischler2019}. There are also residual 3N contributions at third order. Based on our results for the second-order residual term we expect them to be small, but this needs to be confirmed by explicit calculations.
A more detailed study of the zero-temperature MBPT convergence including selected diagrams up to fourth order can be found in Ref.~\cite{Drischler2019}. The convergence behavior of the expansion at finite temperature is similar, with well-converged results for neutron matter being obtained at third order (e.g., for the EMN 450 N$^3$LO interaction the truncation error at third order is at the $100\,\mathrm{keV}$ level at $n=0.2\fmmt$).

Usually we are interested in properties of the EOS at a specific number density $n$. Thus, the relevant thermodynamic potential is the free energy, which is obtained from $\Omega(T,\mu)$ in terms of the Legendre transformation
\begin{align} \label{eq:legendretraf}
    F(T, n) = \Omega(T,\mu) + \mu\, n(T,\mu)\,,
\end{align}
where the number density is given by
\begin{align}
    n(T,\mu) = - \frac{\partial \Omega(T,\mu)}{\partial \mu}\,.
    \label{eq:number-of-particles}
\end{align}
In the $T\rightarrow 0$ limit the free energy gives the ground-state energy of the system.

The free energy determined from Eq.~\eqref{eq:legendretraf} and the perturbation series for $\Omega(T,\mu)$ up to a given order will in general not reproduce the corresponding zero-temperature perturbation series for the ground-state energy. This is because the zero-temperature formalism uses the reference Fermi momentum $k_\text{F}$ whereas grand-canonical MBPT at finite temperature uses the chemical potential $\mu$. In principle one could just use 
the grand-canonical perturbation series also at $T=0$. However, formal arguments and numerical comparisons lead to the conclusion that in general the grand-canonical perturbation series is deficient compared to the zero-temperature one~\cite{Wellenhofer2019}.\footnote{This is particularly evident for a system with a first-order phase transition (like, e.g., symmetric nuclear matter)~\cite{PhysRevC.89.064009}.} 
To obtain a finite-temperature perturbation series that is consistent with the zero-temperature formalism, we follow Kohn and Luttinger~\cite{KohnLuttinger1960} and formally expand the chemical potential as
\begin{align}
    \mu &= \sum_{l=0}^{\infty} \lambda^l \mu_l\label{eq:mu-expansion}\,,
\end{align}
where $\mu_0$ is the chemical potential of the reference system with formally the same density as the interacting system, i.e.,
\begin{align}
    n(T,\mu_0) = - \frac{\partial \Omega_0(T,\mu_0)}{\partial \mu_0}\,.
\label{eq:density}
\end{align}
By inserting the expansion Eq.~\eqref{eq:mu-expansion} into Eq.~\eqref{eq:legendretraf} and reexpanding $\Omega$ and $n$ around $\mu_0$ we obtain
\begin{align}
    F &= \left(\Omega_0^{(0)} - \mu_0 \Omega_0^{(1)}\right)\nonumber + \lambda \Omega_1^{(0)}\\
    &\quad+ \lambda^2 \left( \Omega_2^{(0)} - F_2^a \right)\nonumber\\
    &\quad+ \lambda^3 \left(\Omega_3^{(0)} - F_3^a\right) + \order{\lambda^4}\label{eq:F-expansion}\,,
\end{align}
where $\Omega_l^{(m)} = \partial_\mu^m \eval{\Omega_l \left(T, \mu\right)}_{\mu = \mu_0}$. 
Here, the leading part ${F_0 = (\Omega_0^{(0)} - \mu_0 \Omega_0^{(1)})}$ is the free energy 
of the reference system, and the
additional contributions (due to the expansion about $\mu_0$) at second and third order are given by
\begin{align}
    F_2^a &= \frac{\big(\Omega_1^{(1)}\big)^2}{2 \Omega_0^{(2)}}\label{eq:correction2nd}\,,\\
    F_3^a &= \frac{\Omega_1^{(1)}\Omega_2^{(1)}}{\Omega_0^{(2)}} 
    - \frac{\big(\Omega_1^{(1)}\big)^2 \Omega_1^{(2)}}{2 \big(\Omega_0^{(2)}\big)^2} 
    + \frac{\Omega_0^{(3)} \big(\Omega_1^{(1)}\big)^3}{6 \big(\Omega_0^{(2)}\big)^3}\,.\label{eq:correction3rd}
\end{align}
These expressions are obtained by fixing the higher-order contributions $\mu_i$ in Eq.~\eqref{eq:mu-expansion} 
such that Eq.~\eqref{eq:density} is maintained up to higher-order terms in the expansion of $\mu$ about $\mu_0$.

One can show that for isotropic systems the additional terms given by Eqs.~\eqref{eq:correction2nd}, 
\eqref{eq:correction3rd}, etc., cancel the corresponding anomalous contributions in 
the $T\rightarrow 0$ limit~\cite{KohnLuttinger1960,LuttingerWard1960}.\footnote{The additional terms given by Eqs.~\eqref{eq:correction2nd}, 
\eqref{eq:correction3rd}, etc., have a diagrammatic representation that 
is very similar to the one of anomalous contributions (see, e.g., Refs.~\cite{PhysRevC.89.064009,Wellenhofer2019}).
Note also that these contributions do not vanish individually in the $T\rightarrow 0$ limit, as follows from
$\frac{1}{T} n_\beta \left( 1 - n_\beta \right) = \frac{\partial}{\partial \mu} n_\beta \xrightarrow{T \rightarrow 0} \delta\left(\epsilon_\beta - \mu\right)$.} 
That is,
\begin{align}
\Omega_l^{(0)}(T,\mu_0) - F_l^a(T,\mu_0) \xrightarrow{T\rightarrow 0}  E_l^{(0)}(k_\text{F})\,,
\end{align}
where $E_l^{(0)}(k_\text{F})$ is the contribution of order $l$ in zero-temperature MBPT, 
with $k_\text{F} = (3 \pi^2 n)^{1/3}$ and $n$ the density.
Hence, the reexpanded perturbation series for the free energy, Eq.~\eqref{eq:F-expansion}, 
is consistent with zero-temperature MBPT (in the isotropic case).
In fact, since it does not use the exact chemical potential anymore but only $\mu_0$ whose correspondence to the density 
is to all orders given by Eq.~\eqref{eq:density},
the reexpanded series may be seen to correspond to perturbation theory for the canonical ensemble.

\subsection{Finite-temperature HF-MBPT}
\label{sec22}

\begin{figure}[t]
    \centering
    \includegraphics[width=0.75\columnwidth]{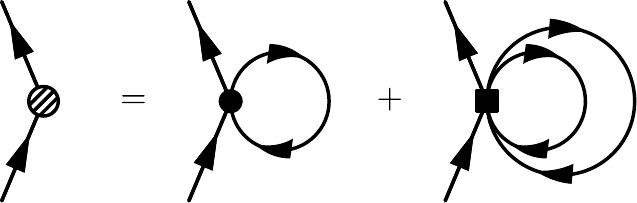}
    \caption{Definition of the single-particle potential $U_\alpha$ (left side) as the self-consistent Hartree-Fock self-energy. Solid dots (squares) represent $V_{\rm NN}$ ($V_{\rm 3N}$) interactions.}
    \label{fig:hf_self_energy}
\end{figure}

The calculations in our paper are carried out using the generalization of HF-MBPT to finite temperatures. Compared to calculations with a noninteracting reference system, using a HF basis is expected to improve the convergence behavior of MBPT~\cite{BOGNER200559,Tichai:2016joa,Drischler2019,Wellenhofer2019}. In zero-temperature and grand-canonical MBPT, respectively, the HF single-particle potential is given by
\begin{align}
U_\alpha(k_\text{F}) &= \frac{\delta E_1(k_\text{F})}{\delta n_\alpha}\,,
\quad\quad
U_\alpha(T,\mu) = \frac{\delta \Omega_1(T,\mu)}{\delta n_\alpha}\,.
\end{align}
Here, the functional derivative 
is defined via
\begin{multline}
\frac{\partial }{\partial \mu} \sum_{\alpha\beta\ldots} f(n_\alpha,n_\beta,\ldots)
= \\
\sum_{\alpha\beta\ldots} \frac{\delta  f(n_\alpha,n_\beta,\ldots)}{\delta n_\xi}  
\frac{\partial n_\xi}{\partial \mu}\bigg|_{\xi\in\{\alpha,\beta,\ldots\}}.
\end{multline}
Explicitly, the expression for the HF potential reads~\cite{HebelerSchwenk2010}
\begin{align}
    U_\alpha &=
    \sum_{\beta} n_\beta \bra{\alpha\beta}\mathcal{A}_{12}V_{\rm NN}\ket{\alpha\beta}\nonumber\\
    &\quad+ \frac{1}{2}\sum_{\beta\gamma} n_\beta n_\gamma \bra{\alpha\beta\gamma}\mathcal{A}_{123}V_{\rm 3N}\ket{\alpha\beta\gamma}\,.
    \label{eq:hf-self-energy}
\end{align}
This matches the expression for the first-order self-energy correction to the in-medium single-particle propagator, as shown diagrammatically in Fig.~\ref{fig:hf_self_energy}. Note that while (for isotropic systems) the evaluation of the HF potential in the zero-temperature formalism is straightforward, in the grand-canonical case it has to be computed self-consistently by solving
\begin{align}  \label{UHFdefinition}
    \varepsilon_\alpha(T,\mu)  = \frac{k^2}{2M} + U_\alpha[T,\mu;\varepsilon_\alpha(T,\mu)]
\end{align}
at fixed $T$ and $\mu$.

\begin{figure}[t]
    \centering
    \includegraphics[width=0.95\columnwidth]{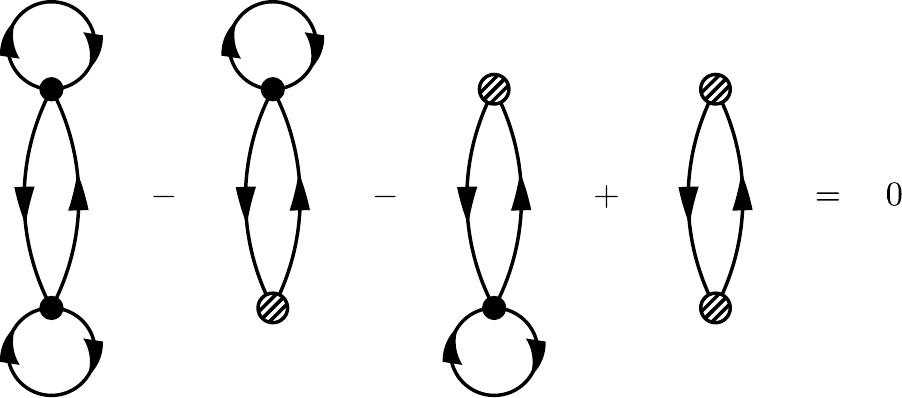}
    \caption{Cancellation of anomalous NN contributions at second order in Hartree-Fock MBPT at finite temperature. The diagrams correspond to Eqs.~\eqref{eq:anomalous-diagram}, \eqref{eq:U1}, \eqref{eq:U2},~and~\eqref{eq:U3} in that order.}
    \label{fig:anourmal_cancelation}
\end{figure}

The key part is now to consistently incorporate the HF potential in the reexpanded perturbation series for the free energy, Eq.~\eqref{eq:F-expansion}. Since it depends on the chemical potential, including the HF potential in the reexpansion would generate additional contributions (via $\mu$ derivatives) that spoil the consistency with zero-temperature MBPT. To rectify this we ``decouple'' the self-consistent HF potential from the thermodynamics by substituting
\begin{align}  \label{UHFtrick1}
    U_\alpha(T,\mu) \longrightarrow U_\alpha(T,\tilde\mu)\,,
\end{align}
where $\tilde\mu$ is an auxiliary ``chemical potential'' (introduced solely as an intermediate tool) that is independent of $\mu$. The reexpansion about $\mu_0$ now leaves the HF potential invariant, and after it is performed we set
\begin{align}  \label{UHFtrick2}
    U_\alpha(T,\tilde\mu) \longrightarrow U_\alpha(T,\mu_0)\,,
\end{align}
leading to the consistent (i.e., canonical) generalization of HF-MBPT to finite $T$.

Apart from improving the convergence of MBPT, 
the self-consistent HF potential
has also the benefit that it removes all contributions associated with 
diagrams that have single-vertex loops. 
In particular, it removes the anomalous contributions as well as 
the additional ones from the reexpansion about $\mu_0$ at second order~\cite{Tolos:2007bh} and third order.
For the second-order anomalous contributions from two-body interactions, this cancellation is depicted diagrammatically in Fig.~\ref{fig:anourmal_cancelation}.
The cancellation occurs because with our choice of $U_\alpha$, the four diagrams of Fig.~\ref{fig:anourmal_cancelation} give matching expressions up to an additional minus sign for the second and third diagram. Furthermore, with our $U_\alpha$ it is $\Omega_1^{(1)} = 0$ after applying Eq.~\eqref{UHFtrick2}, which implies that the correction terms given by
Eqs.~\eqref{eq:correction2nd}~and~\eqref{eq:correction3rd} are zero [since they involve powers of $\Omega_1^{(1)}$]. Hence, with $U_\alpha$ given by the self-consistent HF self-energy (incorporated as described above), 
the canonical perturbation series for the free energy takes the simple form 
\begin{align}
    F(T,\mu_0) &= F_0 + \lambda \Omega_1^{(0)} + \lambda^2 \Omega_{2,\text{normal}}^{(0)} 
    + \lambda^3\Omega_{3,\text{normal}}^{(0)}
    \nonumber \\ & \quad
    + \order{\lambda^4}\,,
    \label{eq:F-HF-expansion}
\end{align}
and the consistency with the zero-temperature formalism is evident.\footnote{Note, however, that new types of anomalous contributions that are not canceled by the HF potential arise at fourth order and beyond.
These would either have to be kept as additional finite-temperature diagrams (together with the corresponding 
terms from the reexpansion about $\mu_0$), or higher-order corrections to the single-particle potential 
would have to be included~\cite{Wellenhofer2019}.}

\section{Results}
\label{sec:results}

In this section we present a systematic study of the neutron matter EOS based on different nuclear interactions obtained from chiral EFT.
First, we employ the NN potentials of Entem, Machleidt, and Nosyk (EMN) \cite{EMN2017} with cutoffs $\Lambda = 450\MeV$ and $\Lambda = 500\MeV$ at orders N$^2$LO and N$^3$LO. Three-nucleon interactions are included up to the same order in the chiral expansion as two-nucleon interactions, using nonlocal regulators with the same cutoff $\Lambda$~\cite{Drischler2019}. Note that the N$^2$LO 3N contributions from the mid- and short-range couplings $c_D$ and $c_E$ vanish in neutron matter for nonlocal regulators~\cite{HebelerSchwenk2010}, and hence our results are independent of the particular $c_D$, $c_E$ fits for all employed interactions in this work.\footnote{Note that three-body contributions proportional to $c_4$ are absent as well in pure neutron matter for all regulators~\cite{HebelerSchwenk2010}.} 
To explore the cutoff dependence we show the variation from $\Lambda = 450\MeV$ to $\Lambda = 500\MeV$ as a band with borders labeled ``EMN N$^2$LO'' or ``EMN N$^3$LO'', respectively. 
These interactions were studied in Ref.~\cite{Drischler2019} up to fourth order in the zero-temperature MBPT expansion, which provides a benchmark for our calculations.

Second, to improve the convergence of the MBPT calculations, we apply the similarity renormalization group (SRG)~\cite{Bogner:2006pc} to decouple low and high momenta via unitary transformations. The resulting low-resolution interactions lead to less correlated wave functions and can lead to a significantly improved convergence of many-body calculations~\cite{Bogner:2009bt}. (Note, however, that unevolved EMN interactions are still sufficiently perturbative to be applicable for the neutron matter calculations presented here.)
In practical calculations the SRG flow cannot be computed exactly but needs to be truncated, typically by discarding all induced operators beyond the three-body level (see, e.g., Refs.~\cite{Jurg09SRG3N,Jurg10SRG3N,Roth11SRG,Roth14SRG3N}). The Hebeler+ interactions of Ref.~\cite{Hebeler_et_al2011} are derived by evolving the N$^3$LO NN potential of Ref.~\cite{EM2003} to resolution scales $\lambda_{\text{SRG}}$, while the 3N interactions at N$^2$LO are determined at the corresponding resolution scale by fits to the $^3$H binding energy and the $^4$He radius using the cutoff $\Lambda_{\rm 3N}$. In Ref.~\cite{Hebeler_et_al2011}, different NN+3N interactions were derived, each characterized by $\lambda_{\text{SRG}}/\Lambda_{\text{3N}}$. In this work we in particular employ the interactions ``1.8/2.0'', ``2.8/2.0'', ``2.0/2.5'' and ``2.0/2.0~(PWA)'', where for the last a different set of long-range 3N couplings has been used (see Ref.~\cite{Hebeler_et_al2011} for details). Finally, we also employ new interactions from Ref.~\cite{Hebeler2020}, where NN+3N interactions are consistently SRG evolved to scales $\lambda_{\text{SRG}}$ using the framework of Ref.~\cite{Hebe12msSRG}.
For all interactions, we include NN partial waves up to total angular momentum $J_{12} \leqslant 8$. Three-nucleon partial waves are included up to $J_\text{tot} \leqslant 9/2$ and $J_{12} \leqslant 5$ or $6$ for SRG-evolved and EMN interactions, respectively. We have checked that these truncations give converged results below the $100$-$\text{keV}$ level.

To test the sensitivity to the SRG resolution scale $\lambda_{\text{SRG}}$, we show the variation from $\lambda_{\text{SRG}} = 1.8\fmmo$ to $\lambda_{\text{SRG}} = 2.8\fmmo$ as bands with borders labeled ``NN~SRG~+~3N~fit'' for the Hebeler+ interactions (with $\Lambda_{\rm 3N} = 2.0 \fmmo$) and ``NN~SRG~+~3N~SRG'' for the consistently evolved interactions. 
The cutoff and SRG scale variations are only one source of uncertainty.
Uncertainty estimates based on the convergence of the EFT expansion are studied in Sec.~\ref{sec:chiral-expansion} and are depicted in Figs.~\ref{plot:F-P-EFT-errors-comparison}~and~\ref{plot:thermal-qtys} as bands without borders.

\begin{figure}[t]
    \centering
    \includegraphics[width=\columnwidth]{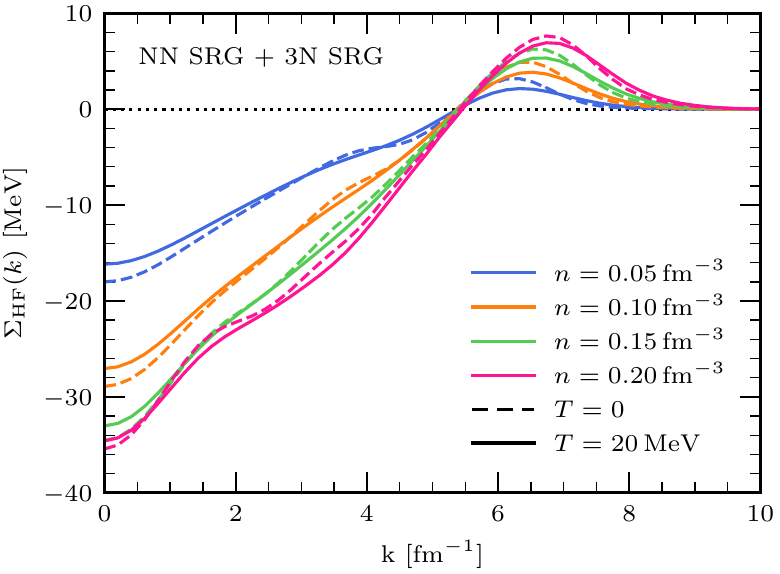}
    \caption{Self-consistent Hartree-Fock self-energy $\Sigma_{\rm HF}(k)$ as a function of momentum $k$ at temperatures ${T=0}$ (dashed) and  $T=20\MeV$ (solid lines) for different densities, 
    obtained from the consistently SRG-evolved NN+3N interaction with $\lambda_{\text{SRG}} = 1.8\,\mathrm{fm}^{-1}$.
    }
    \label{plot:self-energy}
\end{figure}

\begin{figure*}[t]
    \centering
    \includegraphics[width=2\columnwidth]{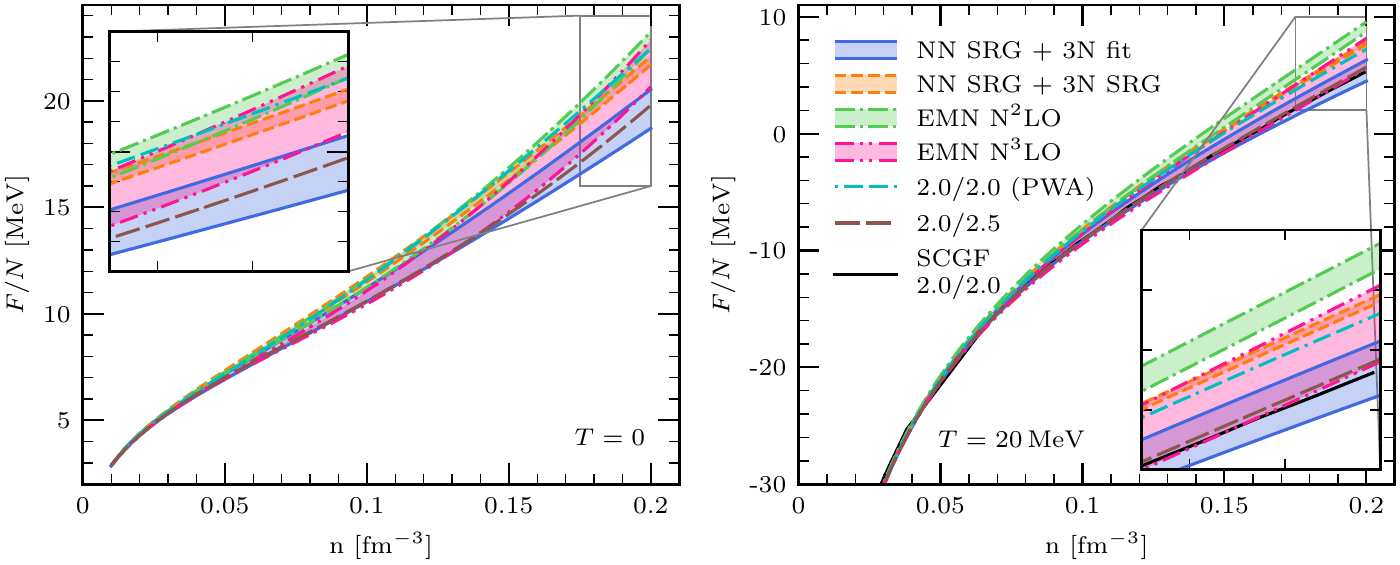}
    \caption{Free energy per particle, $F/N$, as function of the density $n$ at $T=0$ (left) and $T=20\,$MeV (right panel) for different chiral interactions. Bands display variations of the SRG scale from $\lambda_{\text{SRG}} = 1.8\,\mathrm{fm}^{-1}$ to $\lambda_{\text{SRG}} = 2.8\,\mathrm{fm}^{-1}$ for both sets of SRG-evolved interactions, while for the EMN interactions they display the cutoff variations from $\Lambda = 450\,$MeV to $\Lambda = 500\,$MeV. Results for the interactions ``2.0/2.5'' and ``2.0/2.0~(PWA)'' are shown as individual lines. See the text for more details about the interactions. The insets display the density range from $n=0.175\,\mathrm{fm}^{-3}$ to $n=0.2\,\mathrm{fm}^{-3}$.\label{plot:F-T0-T20}}
\end{figure*}

For all results in the following we employ a HF partitioning of the Hamiltonian. Therefore, we first show the HF self-energy $\Sigma_\mathrm{HF}(k)$ in Fig.~\ref{plot:self-energy} for different densities at ${T=0}$ and ${T=20\MeV}$. Here and in the following we use $T=10^{-3}\MeV$ to obtain zero-temperature results with our finite-temperature code. We have checked that using even lower temperatures does not change the results and verified that our $T=10^{-3}\MeV$ results can reproduce zero-temperature results from Ref.~\cite{Drischler2019} very well. The expression for the HF self-energy is given by Eq.~\eqref{eq:hf-self-energy}. Note that Fermi-Dirac distribution functions $n_\beta$ depend on the self-energy, such that a self-consistent solution is necessary, in contrast to zero-temperature calculations. We start with a free spectrum and iterate Eq.~\eqref{UHFdefinition} until convergence is reached. The self-consistent HF self-energy is more conveniently obtained by working at fixed density; i.e., we perform the self-consistent iterations of Eq.~\eqref{UHFdefinition} while adjusting at each iteration step $\mu_0$ (resp.~$\tilde\mu$, see Sec.~\ref{sec22}) to $n$ according to Eq.~\eqref{eq:density}.

The results shown in Fig.~\ref{plot:self-energy} are for the ``NN~SRG~+~3N~SRG'' interaction at $\lambda_{\text{SRG}}=1.8 \fmmo$. We find that the self-energy is mainly attractive up to high momenta around $k\approx 6\fmmo$. Three-particle interactions yield repulsive net contributions for momenta $k \lesssim 5\fmmo$ while the temperature dependence of the results is remarkably small. At the highest density shown ($n = 0.2 \fmmt$), the NN contribution to the self-energy is $-69.1 \MeV$ while 3N contributions yield $33.7 \MeV$ for $k=0$ and $T=0$. For very high momenta ($k \gtrsim 10\fmmo$), the self-energy vanishes due to the employed regulators for the NN and 3N interactions. 

\subsection{Free energy, pressure, and entropy}

The free energy is calculated within MBPT using the formalism discussed in Sec.~\ref{sec2}. We include contributions from NN interactions up to third order, while we neglect 3rd order diagrams involving 3N interactions. Momentum integrals in the individual diagrams are evaluated using the Vegas integration algorithm from Ref.~\cite{Lepage1978} where we take the implementation from Ref.~\cite{Hahn2005} (see also Ref.~\cite{Drischler2019}).

In Fig.~\ref{plot:F-T0-T20} we present results for the free energy per particle for $T=0$ (left) and $T=20\MeV$ (right panel). The different bands (and lines) correspond to different interactions, and the bands result from variations of the interaction cutoff scale and the SRG resolution scale (see legend and the previous section for details). Lines at the borders of bands always represent results for one of the Hamiltonians in that given set. Theoretical uncertainty estimates based on the EFT expansion are provided in Sec.~\ref{sec:chiral-expansion}.

\begin{figure*}[t]
     \includegraphics[width=2\columnwidth]{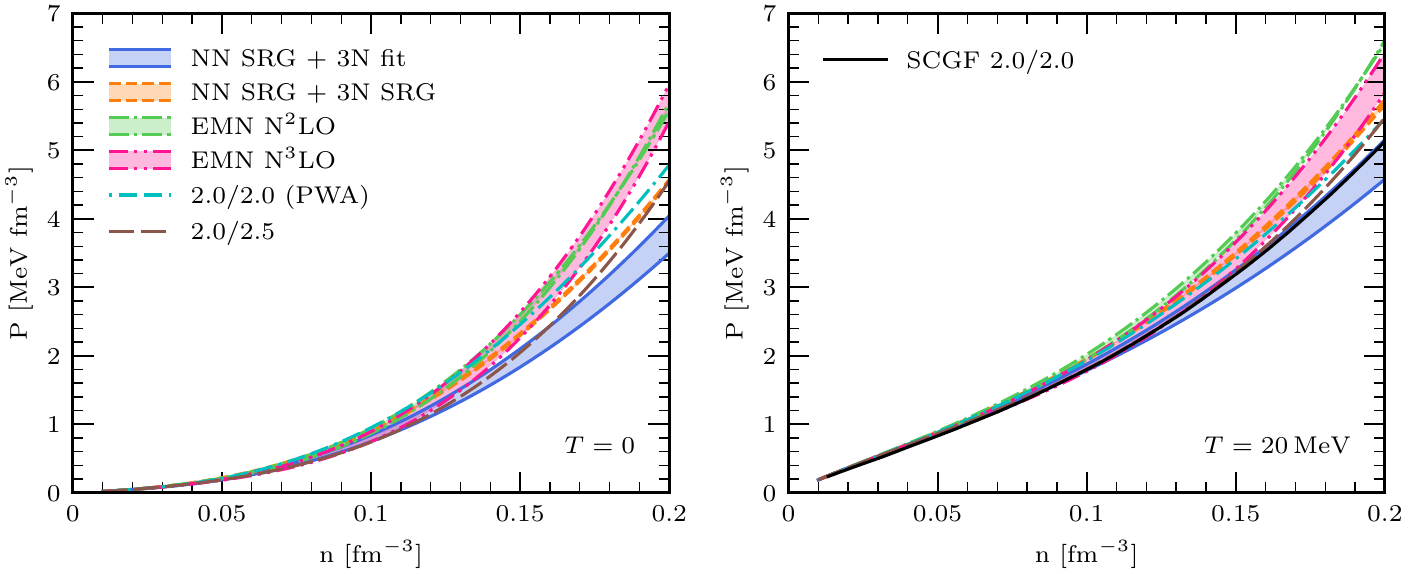}
    \caption{Pressure $P$ as function of the density $n$ at $T=0$ (left) and $T=20\,$MeV (right) for different chiral interactions. For details on how the bands are constructed for the different interactions see the caption of Fig.~\ref{plot:F-T0-T20} and the text. The density derivative $P = n^2 \partial_n (F/N)$ has been calculated analytically by first fitting the results for the interaction free energy
    $F_\mathrm{int}$ via Eq.~\eqref{eq:F_int_fit} while treating the free gas contribution analytically (see text for details).\label{plot:P-T0-T20}}
\end{figure*}

While our results at low densities are almost insensitive to the interactions considered, differences emerge with increasing density. In particular, the size of the cutoff variation bands increases, as expected.
We note that the cutoff dependence of the EMN N$^3$LO interactions is larger than for N$^2$LO in our calculation. This could be due to a slower MBPT convergence at N$^3$LO. Furthermore, we find that the SRG scale dependence of the ``NN~SRG~+~3N~fit'' interactions is comparable to the cutoff sensitivity of the EMN interactions, while the variation of the results for the consistently evolved ``NN~SRG~+3N~SRG'' interactions is much smaller, only about $400\,\mathrm{keV}$ at $n = 0.2\fmmt$ for $T=0$. This indicates that effects from neglected four- and higher-body forces in the SRG evolution are very small for neutron matter in this resolution scale regime (see also Ref.~\cite{Hebeler2020}), and that higher-order MBPT contributions are likely small.

\begin{figure*}[t]
    \centering
    \includegraphics[width=2\columnwidth]{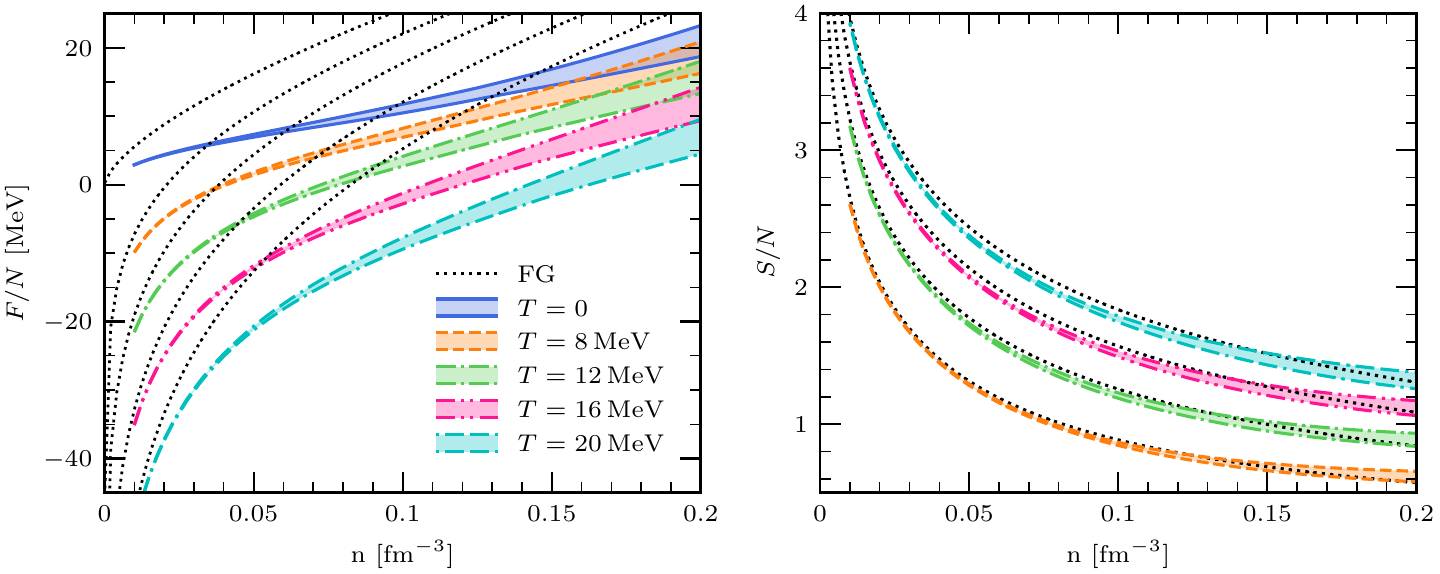}
    \caption{Free energy per particle, $F/N$ (left), and entropy per particle, $S/N$ (right), for $T = 0, 8, 12,16,$ and $20\,\mathrm{MeV}$ as a function of density $n$. The free Fermi gas (FG) is shown as a dotted line. Bands for the free energy are determined by taking the minimum and maximum value of the bands shown in Fig.~\ref{plot:F-T0-T20} and the same procedure is applied for the entropy. Note that the divergence of the free energy per particle for vanishing density originates from the free Fermi gas contribution.\label{plot:F-S-overview}}
\end{figure*}

We determine the pressure
as the density derivative of the free energy, i.e.,
\begin{align}\label{eq:pressure}
    P = n^2 \frac{\partial}{\partial n} \eval{\frac{F}{N}}_T\,.
\end{align}
At finite temperature the free energy per particle diverges logarithmically in the zero-density limit (see, e.g., Ref.~\cite{PhysRevC.89.064009}). This is a result of the free Fermi gas contribution and is also present without interactions. To evaluate Eq.~\eqref{eq:pressure} accurately, we separate the free Fermi gas contribution, which is treated exactly, and differentiate numerically only the interaction free energy,
\begin{align}
F_\text{int}(T, n) = F_{\text{FG}}(T,n) - F(T,n)\,.
\end{align}
(Note that for convenience we define the interaction free energy $F_\text{int}$ as the negative of $F-F_\text{FG}$.) The pressure is then expressed as
\begin{align}
    P(T, n) &= P_{\text{FG}}(T,n) - n^2 \frac{\partial}{\partial n} \eval{\frac{F_\text{int}(T,n)}{N}}_T\,,
\end{align}
where the pressure of the free gas $P_{\text{FG}}(T,n)$ can be evaluated using polylogarithms.
To evaluate the interaction contribution to the pressure, we employ a fit function and calculate the derivative of the fit analytically. We use the function from Ref.~\cite{CarboneSchwenk2019},
\begin{align}
\frac{F_\text{int}(T, n)}{N} = a_0(T) + \sum_{i=1}^4 a_i(T) \left(\frac{n}{n_0}\right)^\frac{i+1}{3}\,,
\label{eq:F_int_fit}
\end{align}
with saturation density $n_0 = 0.16\fmmt$ to set the scale. We also checked that the simpler function $F_\text{int}/N = a \left(n/n_0\right)^\alpha + b \left(n/n_0\right)^\beta$ yields similar results, but with worse fit quality. The results for the pressure are shown in Fig.~\ref{plot:P-T0-T20}. They demonstrate that the model dependence is increased compared to the free energy, as expected for a quantity obtained through a derivative.

To obtain a better insight into the temperature dependence of the EOS, we show the free energy per particle for $T = 0, 8, 12,16,$ and $20\,\mathrm{MeV}$ as a function of density in the left panel of Fig.~\ref{plot:F-S-overview}. For comparison we also show the free energy of the free Fermi gas. Here, for each temperature the respective band combines the individual bands from the different interaction sets shown in Fig.~\ref{plot:F-T0-T20}. The width of the bands increases with increasing density in a comparable way for all temperatures. This reflects the fact that the shift of $F$ for different temperatures is mainly caused by the free Fermi gas contribution; i.e., the temperature dependence of the interaction contribution is small by comparison. The temperature dependence is investigated in more detail in Sec.~\ref{sec:thermal-effects}.

Finally, we calculate the entropy per particle, ${S/N = -\partial_T F/N|_n}$, via
\begin{align}
        S(T, n) = S_{\text{FG}} (T,n) + \frac{\partial}{\partial T} \eval{F_\text{int}(T,n)}_n\,,
\end{align}
where again the free gas contribution is treated analytically and 
the interaction contribution is evaluated by employing a fit function.
The results are shown in the right panel of Fig.~\ref{plot:F-S-overview}. 
The entropy is dominated by the free gas contribution $S_{\text{FG}}$, which is a direct consequence of the weak temperature dependence of $F_\text{int}$ (see also Fig.~\ref{plot:F-int-temperature} and corresponding discussion). As a consequence, the entropy also exhibits only a very weak sensitivity to the employed Hamiltonian.  

\subsection{Chiral expansion}
\label{sec:chiral-expansion}

\begin{figure}[t]
    \centering
    \includegraphics[width=\columnwidth]{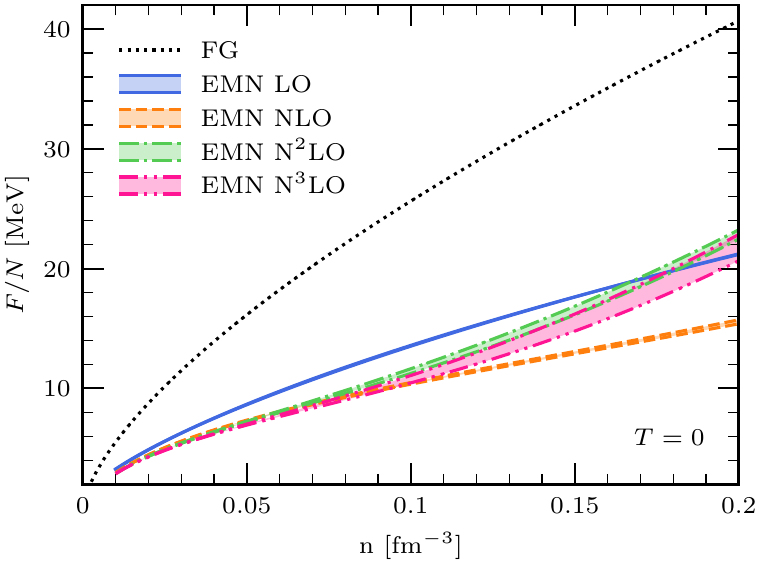}
    \caption{Free energy per particle, $F/N$, for $T=0$ as a function of density $n$ for the EMN interaction at different chiral orders (LO, NLO, N$^2$LO, and N$^3$LO). Bands display the cutoff variation from $\Lambda = 450\,$MeV to $\Lambda = 500\,$MeV. The free Fermi gas (FG) is shown as a dotted line.\label{plot:EFT-order-by-order}}
\end{figure}

Chiral EFT provides a formal expansion in powers of
\begin{align}
    Q = \frac{p}{\Lambda_b}\,,
    \label{eq:chiral_expansion}
\end{align}
where $p$ is the relevant momentum scale for the observable of interest and $\Lambda_b$ the breakdown scale of the EFT. To further investigate the interaction uncertainties, we first show in Fig.~\ref{plot:EFT-order-by-order} the free energy per particle for different orders in the chiral EFT expansion (LO, NLO, N$^2$LO, and N$^3$LO, corresponding to different orders $Q^0$, $Q^2$, $Q^3$, and $Q^4$ in the NN+3N interactions). The narrow bands show the cutoff variation from $\Lambda = 450\MeV$ to $\Lambda = 500\MeV$. The convergence of the chiral expansion is evident in Fig.~\ref{plot:EFT-order-by-order} as the relative contributions get consistently smaller with increasing chiral order. The only exception is the N$^2$LO contribution which is larger than the NLO contribution at densities around $n \approx 0.2 \fmmt$. This is a result of 3N interactions, which start to contribute at N$^2$LO. These give a repulsive contribution which becomes sizable for $n \gtrsim 0.1 \fmmt$ (see the qualitative difference between the NLO and N$^2$LO results shown in Fig.~\ref{plot:EFT-order-by-order}).

\begin{figure}[t]
    \includegraphics[width=\columnwidth]{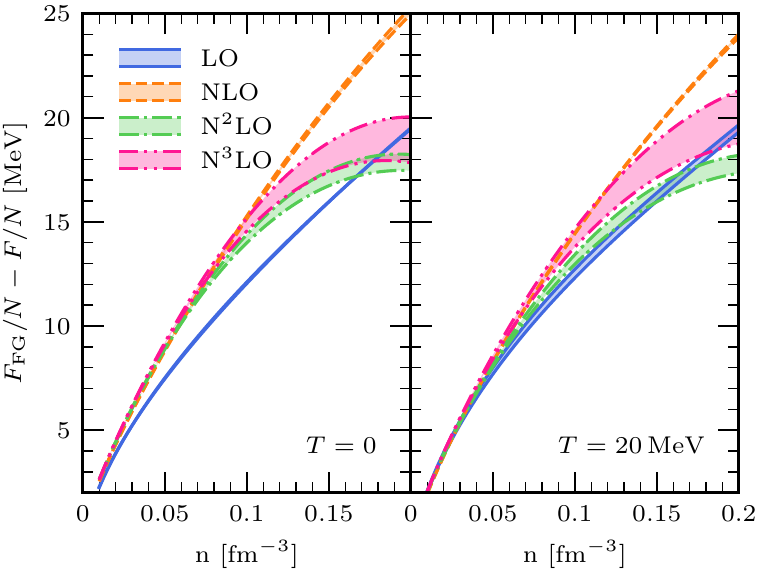}
    \caption{Interaction free energy per particle, $F_\text{int}/N = F_{\text{FG}}/N - F/N$, for $T=0$ (left) and $T=20\,$MeV (right) as a function of density $n$ for the EMN interaction (see Fig.~\ref{plot:EFT-order-by-order} for details and the definition of the bands).\label{plot:F-int-EFT}}
\end{figure}

The convergence of the chiral expansion at finite temperature is examined in Fig.~\ref{plot:F-int-EFT} where we plot the uncertainty bands for the interaction free energy $F_\text{int} = F_{\text{FG}} - F$ as a function of density at the different chiral orders for $T=0$ (left) and $T=20\MeV$ (right). Obviously, the convergence behavior is similar at $T=0$ and at finite temperature. This again reflects the fact that the dominant part of the temperature dependence corresponds to the free gas contribution. For example, the shift $F_\text{int}(T=20\,\text{MeV})-F_\text{int}(T=0)$ is only about $1\MeV$ at $n=0.2\fmmt$, while the shift of the free gas contribution is $-13.8\MeV$ (see also Fig.~\ref{plot:F-int-temperature}). 

A crucial asset of the EFT expansion is the possibility to estimate errors associated with the truncation of the expansion at a finite order. Following Refs.~\cite{EKM2015,PhysRevC.92.024005} we estimate the uncertainty of an observable $X(p)$ in the following way:
\begin{align}
    \Delta X^{(j)} &= Q \cdot \max \Bigl\{ \bigl| X^{(j)} - X^{(j-1)} \bigr|, \Delta X^{(j-1)} \Bigr\}\,,
    \label{eq:eft-uncertainty}
\end{align}
where $X^{(j)}$ denotes the observable calculated from interactions up to order $\mathrm{N}^j\mathrm{LO}$. To apply this prescription to the EOS of neutron matter at a specific density $n$, we follow Ref.~\cite{Drischler2019} and choose the breakdown scale in Eq.~(\ref{eq:chiral_expansion}) equal to $\Lambda_b = 500\,\mathrm{MeV}$ and the momentum scale equal to $p = \sqrt{3/5}\,k_\text{F}$, with $k_\text{F}=(3\pi^2 n)^{1/3}$ the zero-temperature Fermi momentum.

\begin{figure*}[t]
    \centering
    \includegraphics[width=2\columnwidth]{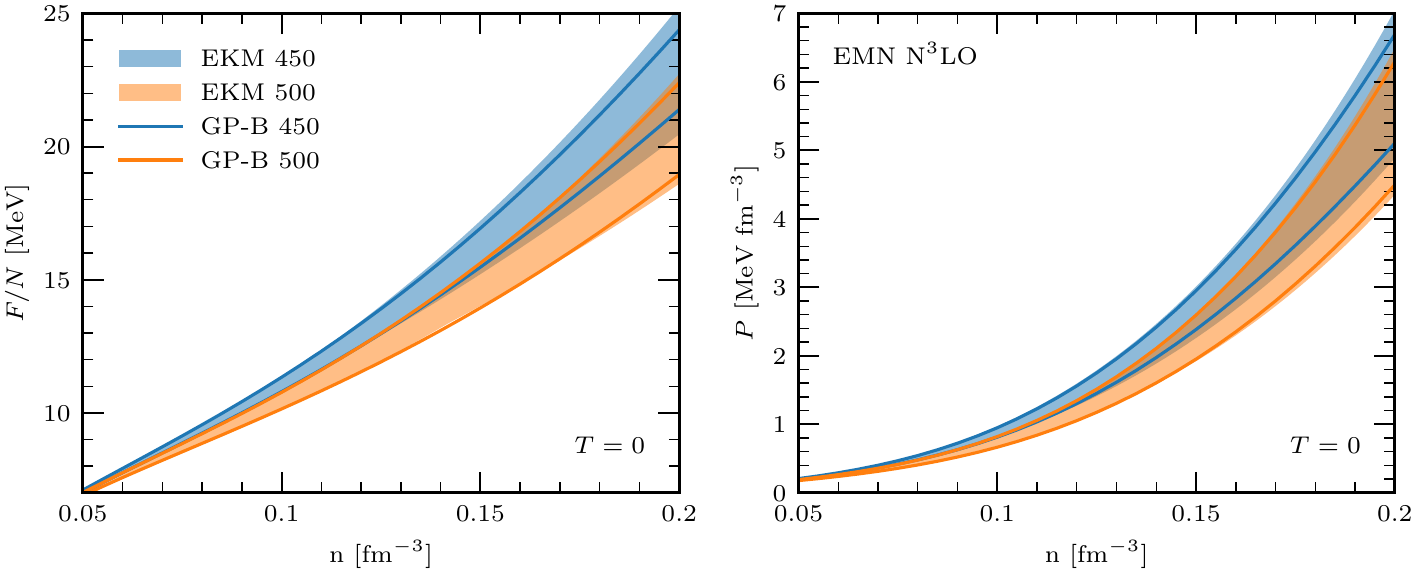}
    \caption{Comparison of the EFT uncertainty estimates calculated by the EKM prescription Eq.~\eqref{eq:eft-uncertainty} (bands) to the Bayesian error estimates based on Gaussian processes (GP-B) of Refs.~\cite{Drischler2020a,Drischler2020b} for the EMN N$^3$LO interaction with ${\Lambda=450\,\mathrm{MeV}}$ and ${\Lambda=500\,\mathrm{MeV}}$. Shown are the free energy per particle, $F/N$ (left), and pressure $P$ (right) as a function of density $n$ for $T=0$. Solid lines mark the boundaries of the $68\%$ GP-B bands.}
    \label{plot:F-P-EFT-errors-comparison}
\end{figure*}

Note that for our uncertainty estimates we omit the leading-order (LO) error. The estimate $\Delta X^\text{LO} = Q^2\abs{X^\text{LO}}$ is problematic in the present context in several ways. First, at nonzero temperature there exists a finite density at which the free energy has a zero crossing, resulting in vanishing errors. Second, at low densities the free energy per particle at finite temperature is dominated by the free gas contribution, and clearly the corresponding enhancement of $\Delta X^\text{LO}$ is unwarranted. These two features could be amended by separating the noninteracting (free Fermi gas) contribution, i.e., by using $\Delta X^\text{LO} = Q^2\abs{X^\text{LO}- X^{\text{FG}}}$ instead. However, we regard it as a clearer strategy to omit the LO error as well as the LO contribution at higher orders in Eq.~\eqref{eq:eft-uncertainty}.

Recently a new Bayesian framework for estimating correlated EFT truncation errors based on Gaussian processes (GP-B) was introduced in Refs.~\cite{Drischler2020a, Drischler2020b}. To provide an alternative error estimate, we apply their publicly available code~\cite{GP-B-code} using $p = k_{\text{F}}$ and $\Lambda_b = 600\,\text{MeV}$ (see Ref.~\cite{Drischler2020a}). A comparison of the prescription by Epelbaum, Krebs, and Mei{\ss}ner (EKM), Eq.~\eqref{eq:eft-uncertainty}, to the \mbox{GP-B} estimate ($68\%$ credible interval) is shown in Fig.~\ref{plot:F-P-EFT-errors-comparison} for the free energy (left) and pressure (right). The EKM prescription Eq.~\eqref{eq:eft-uncertainty} provides slightly larger error estimates, but overall both methods give very similar uncertainty bands.

\subsection{Thermal interaction effects}\label{sec:thermal-effects}

\begin{figure}[t]
    \centering
    \includegraphics[width=\columnwidth]{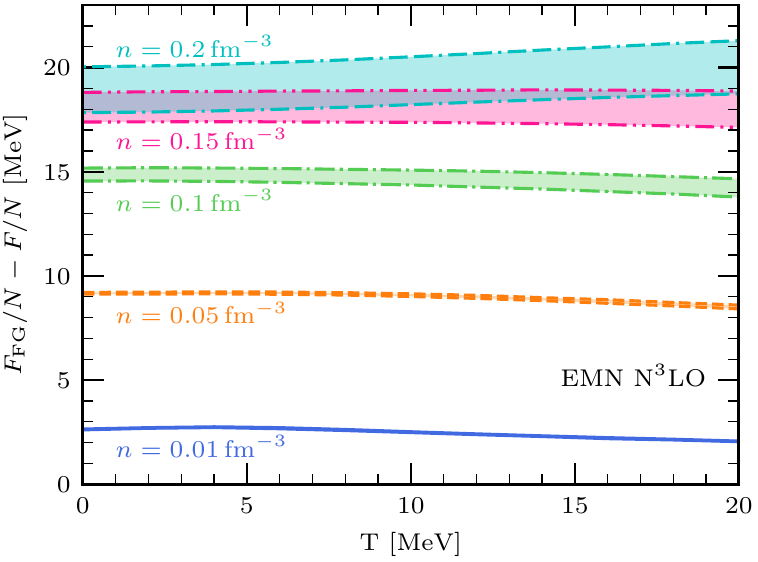}
    \caption{Interaction contribution to the free energy per particle, ${F_\text{int}/N = F_{\text{FG}}/N - F/N}$, as a function of temperature $T$ for different densities $n = 0.01, 0.05, 0.1, 0.15,\,\text{and}~0.2\,\mathrm{fm}^{-3}$, obtained from the EMN N$^3$LO NN+3N interactions (the bands are the same as in Fig.~\ref{plot:F-T0-T20}). The temperature dependence for the other interactions is similarly flat.}
    \label{plot:F-int-temperature}
\end{figure}

Next, we explore thermal effects of the interaction contributions to the EOS. First, in Fig.~\ref{plot:F-int-temperature} we examine the interaction free energy ${F_\text{int} = F_{\text{FG}} - F}$ as a function of temperature for different densities. The results show that the temperature dependence of $F_\text{int}$ is very small for all considered densities, as noted above. To characterize thermal interaction effects in more detail we define the thermal part of a given thermodynamic quantity $X(T,n)$ as the difference between finite-temperature and zero-temperature value, i.e.,
\begin{align}
    X_\text{th}(T, n) &= X(T, n) - X(T=0, n)\,.
\end{align}
From the thermal components of the pressure and internal energy density one obtains a very useful quantity that characterizes thermal effects, the so-called thermal index $\Gamma_\mathrm{th}$:
\begin{align}
    \Gamma_\mathrm{th}(T, n) &= 1 + \frac{P_\mathrm{th}(T, n)}{\mathcal{E}_\mathrm{th}(T, n)}\,,
\end{align}
where $\mathcal{E}_\mathrm{th}=E_\mathrm{th}/V$ is the thermal energy density. The free Fermi gas has $\Gamma_{\mathrm{FG,th}} = 5/3$ independent of density and temperature. Any deviations of $\Gamma_\mathrm{th}(T, n)$ from $5/3$ is thus due to thermal interaction effects. The thermal index is often used to parametrize the temperature dependence of nuclear EOS used in astrophysical simulations~\cite{Mignone:2007fw,Bauswein:2010dn}, where a constant $\Gamma_\mathrm{th}$ independent of $T$ and $n$ (e.g., $\Gamma_{\mathrm{th}} = 1-2$) is sometimes adopted.

\begin{figure}[t!]
    \centering
    \includegraphics[width=\columnwidth]{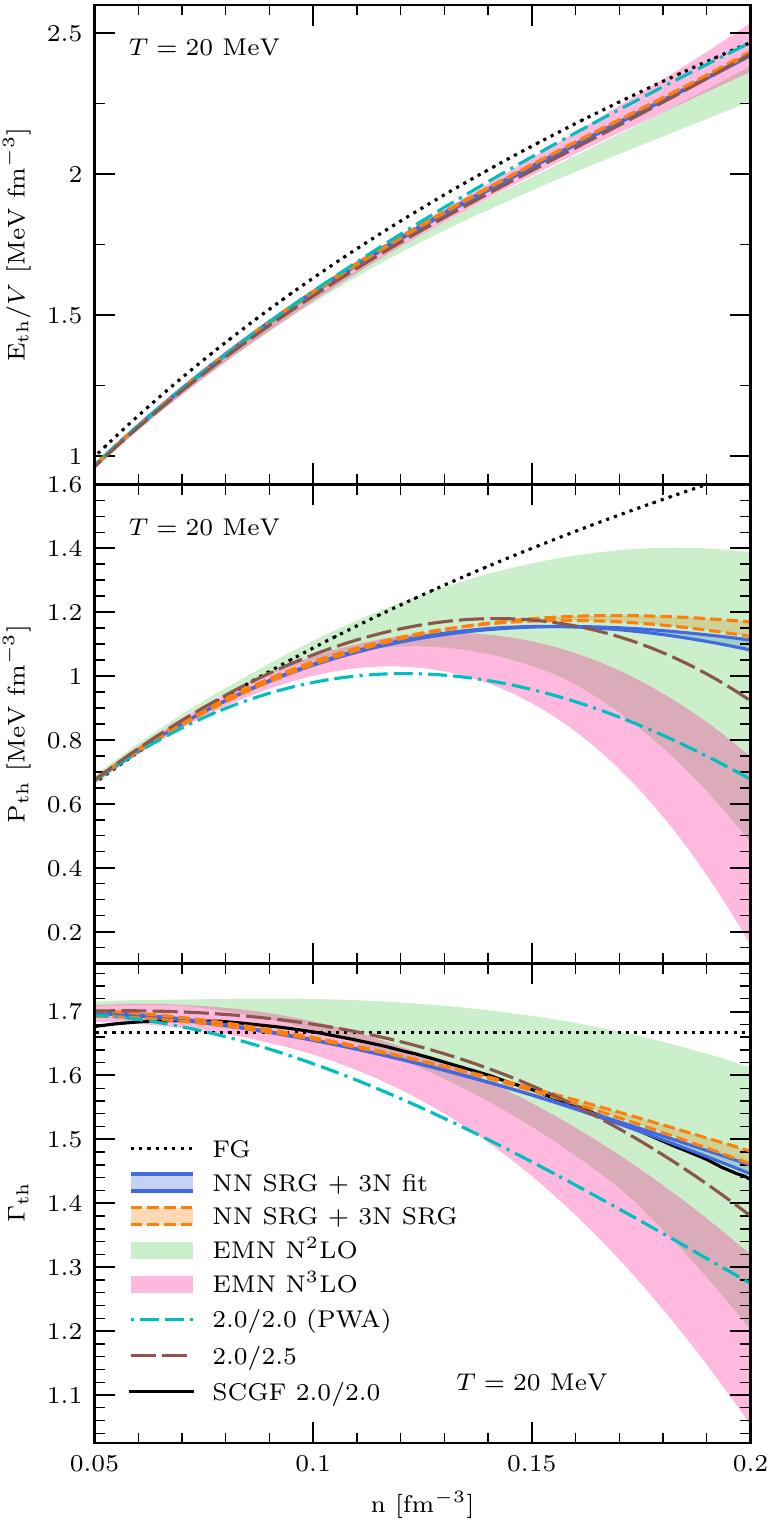}
    \caption{Thermal energy $E_{\rm th}$ (top), thermal pressure $P_{\rm th}$ (middle), and thermal index $\Gamma_{\rm th}$ (bottom) at $T = 20\,\mathrm{MeV}$ as a function of density $n$. Bands with borders are the same as in Fig.~\ref{plot:F-T0-T20}, while bands without borders display EFT uncertainty estimates from the EKM prescription Eq.~\eqref{eq:eft-uncertainty}. The free Fermi gas (FG) is shown as a dotted line. For comparison the thermal index of the ``2.0/2.0'' interaction obtained from self-consistent Green`s function (SCGF) calculations from Ref.~\cite{CarboneSchwenk2019} is also shown (black solid line). The ``2.0/2.0'' interaction is contained in the NN~SRG~+~3N~fit band.}
    \label{plot:thermal-qtys}
\end{figure}

Our results for the thermal energy (top), the thermal pressure (middle) and the thermal index (bottom) at $T=20\MeV$ are displayed in Fig.~\ref{plot:thermal-qtys}. For comparison, we also show as a black solid line the thermal index obtained in Ref.~\cite{CarboneSchwenk2019} using the self-consistent Green`s function (SCGF) approach with the ``2.0/2.0'' interaction of Ref.~\cite{Hebeler_et_al2011}. Our MBPT calculations are consistent with these nonperturbative SCGF results as the SCGF line is very similar to the NN~SRG~+~3N~fit band, which includes the same interaction.

Compared to the thermal pressure and the thermal index, the thermal energy exhibits a much smaller interaction dependence. This can be understood in terms of the decomposition (at fixed density)
\begin{align}
    \Eth(T)
    &= \left(F_\mathrm{FG}(T) - F_\mathrm{FG}(T=0)\right)\\
    &\quad- \left(F_\mathrm{int}(T) - F_\mathrm{int}(T=0) \right) + T \, S(T) \nonumber\,.
\end{align}
Here, $F_\mathrm{int}(T) - F_\mathrm{int}(T=0)$ is small (see Fig.~\ref{plot:F-int-temperature}), and the entropy $S$ deviates only slightly from its free Fermi gas value (see Fig.~\ref{plot:F-S-overview}). Hence, the thermal energy $\Eth$ is dominated by the free gas contribution. The thermal pressure and the thermal index, however, involve the density derivative of $F_\mathrm{int}(T,n)$ and thus deviate more significantly from the corresponding Fermi gas values and, as a consequence, have larger uncertainties. In particular, 3N interactions have a crucial effect on their density dependence. The thermal pressure would increase with increasing density if 3N interactions were not included, as found also in Ref.~\cite{CarboneSchwenk2019}.

\begin{figure}
    \centering
    \includegraphics[width=\columnwidth]{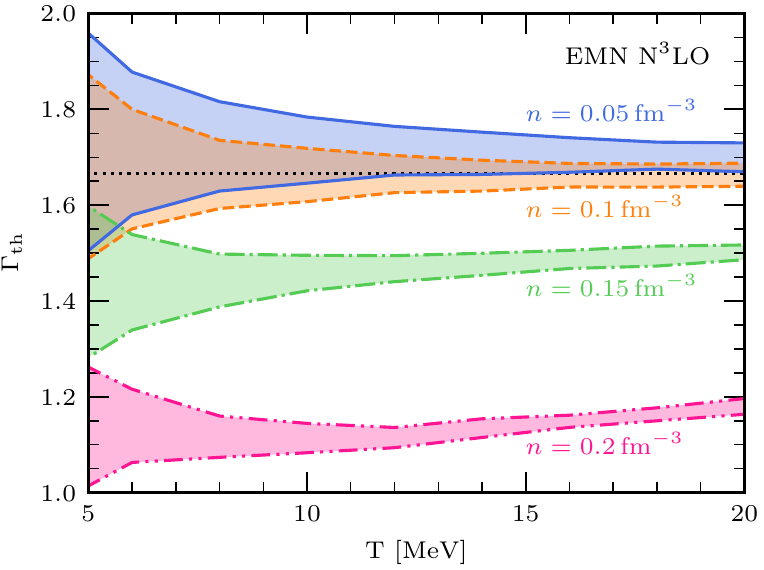}
    \caption{Temperature dependence of the thermal index $\Gamma_{\rm th}$ of the EMN N$^3$LO interaction for $n = 0.05, 0.1, 0.15$, and $0.2\,\mathrm{fm}^{-3}$. The bands combine cutoff variation from $\Lambda=450\,\mathrm{MeV}$ to $\Lambda=500\,\mathrm{MeV}$ with a constant error estimate for the Monte Carlo integration (see text for details). The dotted line marks $\Gamma_{\mathrm{FG,th}} = 5/3$.}
    \label{plot:G-th-temperature}
\end{figure}
 
The temperature dependence of the thermal index $\Gth$ is shown in Fig.~\ref{plot:G-th-temperature} for the EMN N$^3$LO interaction. Since the index is defined as $\Gth=1+P_\mathrm{th}/\mathcal{E}_\mathrm{th}$, the thermal index is very sensitive to uncertainties in $P_\mathrm{th}$ and $E_\mathrm{th}$ at low temperatures (and low densities) where both these quantities are small. Therefore, in addition to the cutoff-variation band we include in Fig.~\ref{plot:G-th-temperature} also an estimate of the numerical Monte Carlo integration errors for these quantities, where we have chosen $\Delta\Pth = 20\,\mathrm{keV \, fm}^{-3}$ and $\Delta\Eth/N = 20\,\mathrm{keV}$. As seen in Fig.~\ref{plot:G-th-temperature}, this leads to sizable uncertainties for $\Gth$ at low temperatures. For $T \gtrsim 10\MeV$ the uncertainties are better controlled and we see only a weak temperature dependence of the thermal index. This behavior is similar for all the other interactions considered.

\subsection{Effective mass approximation}

In the previous section we showed that the thermal index $\Gth(T,n)$ exhibits only a very weak temperature dependence (see Fig.~\ref{plot:G-th-temperature}). Here, we now make use of this feature to construct an approximate parametrization of thermal effects in terms of a density-dependent effective neutron mass $m_n^*(n)$. The thermal index $\Gamma^*_\text{th}(n)$ of an ideal gas of fermions with density-dependent effective mass $m_n^*(n)$ can be expressed as (see, e.g., Ref.~\cite{Constantinos_et_al_2015})
\begin{align}
	\Gamma^*_\text{th}(n) =  \frac{5}{3} - \frac{n}{m_n^*} \frac{\partial m_n^*}{\partial n}\,.
     \label{eq:free-gas-effective-mass}
\end{align}

\begin{figure}[t!]
    \centering
    \includegraphics[width=\columnwidth]{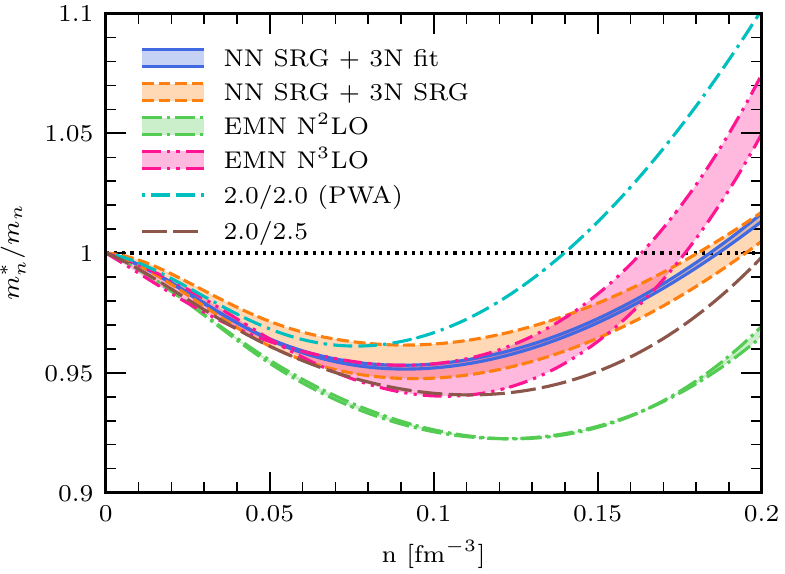}
    \caption{Results for the neutron effective mass $m_n^*(n)$ as a function of density $n$ derived from the thermal index $\Gth(T,n)$ at $T=20\MeV$ as discussed in the text. Results are shown for different NN+3N interactions (same as in Fig.~\ref{plot:F-T0-T20}).}
    \label{plot:meff}
\end{figure}

In Ref.~\cite{CarboneSchwenk2019} it was demonstrated that $\Gamma^*_\text{th}$ determined via Eq.~\eqref{eq:free-gas-effective-mass}, with an effective mass taken from microscopic calculations, agrees well with the thermal index determined by $\Gth = 1 + \Pth / \mathcal{E}_\mathrm{th}$. That means, by taking for $\Gamma^*_\text{th}(n)$ our microscopic results for $\Gth(T,n)$ at $T=20\MeV$ shown in Fig.~\ref{plot:thermal-qtys}, we can integrate Eq.~\eqref{eq:free-gas-effective-mass} to obtain\footnote{Note that uncertainties of $\Gth$ are enhanced at low densities (see Fig.~\ref{plot:G-th-temperature} and discussion) so that $m^*$ obtained by integrating Eq.~\eqref{eq:free-gas-effective-mass} is an approximation.} $m_n^*(n)$. For this we use $m_n^*/m_n(n=0) = 1$, $\Gamma_\mathrm{th}(n=0) = 5/3$, and interpolate linearly to our lowest-density result for $\Gth$ at $n=0.01\fmmt$.

The results for the neutron effective mass $m_n^*(n)$ determined by this procedure are shown in 
Fig.~\ref{plot:meff}. 
The bands display cutoff or SRG scale variations (see caption of Fig.~\ref{plot:F-T0-T20}). We observe that $m_n^*(n)$ first decreases with increasing density, while at around $n\gtrsim 0.1\fmmt$ the effective mass starts to increase again. This effect is related to the contribution of 3N interactions. Based only on NN interactions, the resulting effective mass would decrease with density. A similar qualitative behavior is also found in the SCGF calculations of Ref.~\cite{CarboneSchwenk2019}.

From our results for the effective mass $m_n^*(n)$ we can construct an approximate parametrization of the temperature dependence of the EOS. For this, we again separate thermodynamic quantities into cold and thermal parts, e.g., for the pressure
\begin{align}
    P(T,n) = P(T=0,n) + \Pth(T,n)\,.
\end{align}
The thermal part $\Pth(T)$ is now approximated by
\begin{align}
    \Pth(T,n) \approx P_{\mathrm{FG}, \mathrm{th}}^{m^*}(T, n)\,,
\end{align}
where $P_{\mathrm{FG}, \mathrm{th}}^{m^*}$ is the thermal pressure of an ideal gas of neutrons with density-dependent mass $m_n^*(n)$, i.e., 
\begin{align}
     P_{\mathrm{FG}, \mathrm{th}}^{m^*}(T, n) = n^2 \frac{\partial}{\partial n} \frac{F_{\mathrm{FG},\mathrm{th}}(T, n, m_n^*(n))}{N}\,,
\end{align}
where $F_{\mathrm{FG},\mathrm{th}}(T, n, m_n^*(n))$ is the expression for the thermal free energy of the free neutron gas with $m$ substituted by $m_n^*(n)$.

\begin{figure}[t]
    \centering
    \includegraphics[width=\columnwidth]{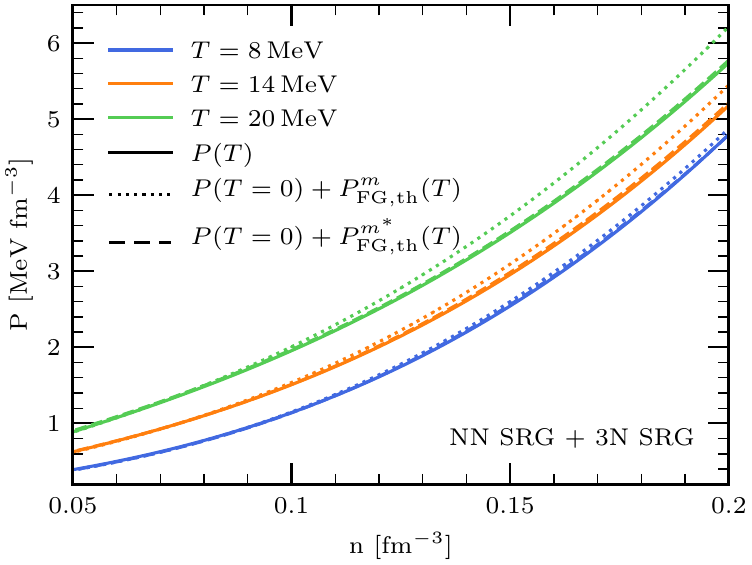}
    \caption{Comparison of the full temperature-dependent pressure (solid lines) to ideal Fermi gas approximations of the thermal contribution with bare neutron mass $P_{\mathrm{FG}, \mathrm{th}}^m(T)$ (dotted lines) and with effective neutron mass $P_{\mathrm{FG}, \mathrm{th}}^{m^*}(T)$ from Fig.~\ref{plot:meff} (dashed lines). The employed effective mass $m_n^*(n)$ is extracted at $T=20\,\mathrm{MeV}$. See text for more details. Shown are temperatures $T=8,14$, and $20\,\mathrm{MeV}$ (different colors) for the consistently SRG-evolved NN+3N interaction with $\lambda_{\text{SRG}} = 1.8\,\mathrm{fm}^{-1}$.
    The dashed lines overlap excellently with the respective solid lines for $T=8\,\mathrm{MeV}$ and $T=14\,\mathrm{MeV}$.}
    \label{plot:P-meff}
\end{figure}

With the microscopic calculations at zero and finite temperature at hand, we now investigate the quality of such an approximation. That is, we compare the results for the pressure $P(T,n)$ obtained using three different ways to calculate its thermal part $\Pth$:
\begin{enumerate}
\item the full finite-temperature calculation for $\Pth(T,n)$, 
\item the ideal gas approximation with bare neutron mass $\Pth(T,n)\approx P_{\mathrm{FG}, \mathrm{th}}^{m}(T,n)$, and
\item the ideal gas approximation with density-dependent effective mass $\Pth(T,n)\approx P_{\mathrm{FG}, \mathrm{th}}^{m^*}(T,n)$.
\end{enumerate}
The results are shown in Fig.~\ref{plot:P-meff}. The effective-mass approximation $P_{\mathrm{FG}, \mathrm{th}}^{m^*}$ reproduces excellently the full finite-temperature calculation $\Pth$, whereas results based on the bare mass $P_{\mathrm{FG}, \mathrm{th}}^{m}$ deviate from the full finite-temperature calculation, with an increasing error as the density increases. This demonstrates that $\Gth(T,n) \approx \Gth^*(n)$ and $m^*_n(n)$ capture the 
finite-temperature effects of the neutron matter EOS very well.

\section{Conclusion and outlook}
\label{sec:conclusions}

In this paper, we studied the neutron matter EOS at finite temperature using MBPT. After discussing the many-body formalism for a general partitioning of the Hamiltonian and the anomalous diagrams at finite temperature, we showed how the many-body expansion simplifies when using a HF partitioning and performed calculations in this scheme. In contrast to previous finite-temperature MBPT studies, we included the full HF self-energy momentum dependence and do not employ normal-ordering approximations for the 3N interactions. For the practical calculations we employed Monte Carlo integration techniques that allow to evaluate highly dimensional integrals very efficiently and make it possible to include all contributions from NN interactions completely up to third order in the many-body expansion and contributions from 3N interactions up to second order including residual contributions.

We then presented a systematic study of the thermodynamics of neutron matter based on a range of chiral EFT interactions. This included the Hebeler+ potentials from Ref.~\cite{Hebeler_et_al2011} as well as for the first time consistently SRG-evolved interactions~\cite{Hebeler2020} in nuclear matter calculations. In addition, we studied the EMN potentials at N$^2$LO and N$^3$LO~\cite{EMN2017} supplemented with 3N interactions used for nuclear matter and nuclei in Refs.~\cite{Drischler2019,Hoppe:2019uyw}. Our results based on the consistently SRG-evolved interactions exhibit a remarkably small SRG scale dependence over the full range of temperatures, which indicates that the effects of induced higher-body forces are very small for these interactions and also that the many-body calculation is well converged. In addition, we studied the theoretical uncertainties due to the truncation of the chiral expansion using the EKM prescription~\cite{EKM2015} and employing the recently developed Bayesian framework based on Gaussian processes~\cite{Drischler2020a,Drischler2020b}. Our results show that both methods provide very similar error estimates. 

Finally, the temperature dependence of different thermodynamic quantities was studied in detail.
We found that the dominant contribution to the temperature dependence originates from the Fermi gas contribution, and that the thermal interaction part is well captured by using a density-dependent effective mass. This was shown by studying the thermal index, which allows to diagnose in a simple way thermal interaction effects.

The present work lays the foundation for microscopic studies of the thermodynamics of isospin-asymmetric nuclear matter based on modern NN and 3N interactions up to high orders in the chiral expansion. The framework allows to incorporate any nuclear interactions that are available in a partial-wave decomposed form and makes it possible to extend the many-body calculations in a transparent way by including higher-order terms, which might be necessary at nonzero proton fractions. Finally, it will be interesting to explore the resulting EOSs at finite temperatures in astrophysical simulations of core-collapse supernovae and neutron star mergers.

\begin{acknowledgments}
This work is supported in part by the Deutsche Forschungsgemeinschaft (DFG, German Research Foundation) -- Project-Id 279384907 -- SFB 1245.
\end{acknowledgments}

\bibliographystyle{apsrev4-2}
\bibliography{paper}

\end{document}